\documentclass[10pt,a4paper]{article}
\usepackage{amsmath,amsfonts,amssymb,amscd}
\catcode`\@=11
\@addtoreset{equation}{section}

\catcode`\@=12

\def\de#1/de#2{\frac{\partial {#1}}{\partial {#2}}}

\begin{document}
%%%%%%%%%%%%%%%%%%%%%%%%%%%%%%%%%%%%%%%%%%%%%%%%%%%%%%%%%%%%%%%%%%%%%%%%%%%%%%%%%%%%%%%%%%%%%%%%%%%
\title{\textbf{A modified theory of gravity with torsion\\ and its applications to cosmology\\ and particle physics}}
\author{Luca Fabbri$^{1,2}$\footnote{E-mail: fabbri@diptem.unige.it}
\footnote{E-mail: luca.fabbri@bo.infn.it} \ and 
Stefano Vignolo$^{1}$\footnote{E-mail: vignolo@diptem.unige.it}\\
\footnotesize{$^{1}$DIME Sez. Metodi e Modelli Matematici, Universit\`{a} di Genova}\\
\footnotesize{Piazzale Kennedy Pad. D, 16129 Genova, Italy}\\
\footnotesize{$^{2}$INFN \& Dipartimento di Fisica, Universit\`{a} di Bologna}\\
\footnotesize{Via Irnerio 46, 40126 Bologna, Italy}}
\date{}
%%%%%%%%%%%%%%%%%%%%%%%%%%%%%%%%%%%%%%%%%%%%%%%%%%%%%%%%%%%%%%%%%%%%%%%%%%%%%%%%%%%%%%%%%%%%%%%%%%%
\maketitle
%%%%%%%%%%%%%%%%%%%%%%%%%%%%%%%%%%%%%%%%%%%%%%%%%%%%%%%%%%%%%%%%%%%%%%%%%%%%%%%%%%%%%%%%%%%%%%%%%%%
\begin{abstract}
\noindent In this paper we consider the most general least-order derivative theory of gravity in which not only curvature but also torsion is explicitly present in the Lagrangian, and where all independent fields have their own coupling constant: we will apply this theory to the case of ELKO fields, which is the acronym of the German \textit{Eigenspinoren des LadungsKonjugationsOperators} designating eigenspinors of the charge conjugation operator, and thus they are a Majorana-like special type of spinors; and to the Dirac fields, the most general type of spinors. We shall see that because torsion has a coupling constant that is still undetermined, the ELKO and Dirac field equations are endowed with self-interactions whose coupling constant is undetermined: we discuss different applications according to the value of the coupling constants and the different properties that consequently follow. We highlight that in this approach, the ELKO and Dirac field's self-interactions depend on the coupling constant as a parameter that may even make these non-linearities manifest at subatomic scales.
\end{abstract}
%%%%%%%%%%%%%%%%%%%%%%%%%%%%%%%%%%%%%%%%%%%%%%%%%%%%%%%%%%%%%%%%%%%%%%%%%%%%%%%%%%%%%%%%%%%%%%%%%%%
\section{Introduction}
The debate in general relativity about the presence of torsion, although one of the less known, is nevertheless one of the most fascinating, and several reasons have been put on the table, both against and in favour of torsion, ranging from philosophical to mathematical and phenomenological aspects of science, covering the foundational, theoretical and applicative domains of physics: the very first of these reasons is related to the fact that once the covariant derivative is written in the most general way, it is defined upon a connection that in general is not symmetric in the two lower indices, and therefore torsion does not generally vanish; the inclusion of torsion beside the curvature makes it possible to introduce a torsion-spin coupling beside the curvature-energy coupling, fully realizing the geometry-matter coupling prescription, which is lame otherwise; if we focus either on the standard or on more exotic spinor fields, the torsional effects manifest themselves as spinorial self-interactions. Nevertheless, a problem that still needs to be addressed is that torsional effects appear to affect the dynamics only at the Planck scale, which would make torsion nearly negligible in almost every practical situation; this circumstance is due to the fact that, since both torsion and curvature have the same geometrical origin, then they are supposed to have the very same coupling constant, and a torsion having the gravitational coupling constant is what makes torsion interact weakly except at the Planck scale.

On the other hand, despite torsion and curvature come in fact from the geometrical background, nevertheless they are independent fields, and therefore they should possess two independent coupling constants; previously \cite{f-v} we have proposed a least-order derivative generalization of the Hilbert-Einstein-Cartan-Sciama-Kibble HECSK Lagrangian in which a quadratic torsional term is present, with its own coupling. We have eventually studied an application to the case of the Dirac field. We have shown that when properly tuned there are consequences that could be detectable at subatomic scales.

However, since torsion has three irreducible decompositions, then each of these should be thought as an independent field, entering with its own coupling constant; in this paper, we shall propound the most general least-order derivative generalization of the HECSK Lagrangian in which one quadratic torsional term for each of the three irreducible parts of torsion is present, and these three quadratic torsional terms will have three coupling constants. We will study applications to two cases: the case of ELKO spinors, whose name means \textit{Eigenspinoren des LadungsKonjugationsOperators}, which designates eigenspinors of the charge conjugation operator, so that they are a Majorana-like spinor for which the problem of the mass term is solved by employing second-order derivative scalar-like field equations; the case of Dirac spinors, the most general type of spinor fields with least-order derivative field equations. In both cases the choice of finely-tuned couplings will have consequences stretching from cosmology, where we will study the case of ELKO fields, to particle physics, in which we will discuss the case of Dirac fields.

The paper is organized as in the following: in the first section, we shall discuss why we propose a generalized gravitational Lagrangian for curvature and torsion, all independent fields having its own coupling constant, writing down the most general Lagrangian and the corresponding field equations; in the second section we shall apply this model to both ELKO and Dirac fields, showing how torsion endows ELKO and Dirac field equations with self-interactions having free coupling constants.
%%%%%%%%%%%%%%%%%%%%%%%%%%%%%%%%%%%%%%%%%%%%%%%%%%%%%%%%%%%%%%%%%%%%%%%%%%%%%%%%%%%%%%%%%%%%%%%%%%%
\section{The geometrical structure of the field equations}
In this paper, we consider spacetime $(1+3)$-dimensional manifolds possessing curvature and torsion, with its three irreducible decompositions, accounting for a total of four independent geometrical fields, given by the metric tensor, the torsion completely antisymmetric dual of an axial vector, the torsion trace vector and the torsion non-completely antisymmetric irreducible tensor; the metric tensor is given by $g_{\alpha\sigma}$ with the covariant derivative denoted by $D_{\mu}$ defined through the dynamical connection $\Gamma^{\mu}_{\alpha\sigma}$: metricity $Dg=0$ holds.

In terms of the connection alone it is also possible to define two additional structures that are fundamental when studying the non-commutative properties of the covariant derivatives, one of which is the Riemann curvature tensor defined as usual by 
\begin{eqnarray}
G^{\rho}_{\phantom{\rho}\eta\mu\nu}
=\partial_{\mu}\Gamma^{\rho}_{\eta\nu}-\partial_{\nu}\Gamma^{\rho}_{\eta\mu}
+\Gamma^{\rho}_{\sigma\mu}\Gamma^{\sigma}_{\eta\nu}
-\Gamma^{\rho}_{\sigma\nu}\Gamma^{\sigma}_{\eta\mu}
\end{eqnarray}
while the other is the Cartan torsion tensor defined as the antisymmetric part in the connection as in the following
\begin{eqnarray}
Q^{\rho}_{\phantom{\rho}\mu\nu}
=\Gamma^{\rho}_{\mu\nu}-\Gamma^{\rho}_{\nu\mu}
\end{eqnarray}
so that once the metric is considered, also the contorsion tensor can be defined as
\begin{eqnarray}
K^{\rho}_{\phantom{\rho}\mu\nu}
=\frac{1}{2}\left(Q^{\rho}_{\phantom{\rho}\mu\nu}
+Q_{\mu\nu}^{\phantom{\mu\nu}\rho}+Q_{\nu\mu}^{\phantom{\nu\mu}\rho}\right)
\end{eqnarray}
and because of their definition together with the metricity condition $Dg=0$, we have that the curvature $G_{\sigma\eta\rho\nu}$ is antisymmetric in both the first and second couple of indices, while torsion $Q_{\rho\mu\nu}$ is antisymmetric in the last two indices and contorsion $K_{\rho\mu\nu}$ is antisymmetric in the first two indices; due to these antisymmetry properties, the curvature tensor has one independent contraction given by $G^{\rho}_{\phantom{\rho}\eta\rho\nu}=G_{\eta\nu}$ with contraction $G_{\eta\nu}g^{\eta\nu}=G$ as usual, while both torsion and contorsion have one independent contraction that is given by $K_{\nu\rho}^{\phantom{\rho\nu}\rho}= Q^{\rho}_{\phantom{\rho}\rho\nu}=V_{\nu}$: with trace vector $V_{\nu}$ and completely antisymmetric dual of the axial vector $2K_{\rho\mu\nu}\varepsilon^{\rho\mu\nu\alpha}= Q_{\rho\mu\nu}\varepsilon^{\rho\mu\nu\alpha}=W^{\alpha}$, it is possible to see that the most general torsion can actually be decomposed in irreducible decompositions according to the identities
\begin{eqnarray}
\nonumber
&Q_{\rho\mu\nu}
=\frac{1}{6}W^{\alpha}\varepsilon_{\alpha\rho\mu\nu}
+\frac{1}{3}\left(V_{\nu}g_{\rho\mu}-V_{\mu}g_{\rho\nu}\right)+\\
&+[Q_{\rho\mu\nu}-\frac{1}{6}W^{\alpha}\varepsilon_{\alpha\rho\mu\nu}
-\frac{1}{3}\left(V_{\nu}g_{\rho\mu}-V_{\mu}g_{\rho\nu}\right)]
\end{eqnarray}
and also
\begin{eqnarray}
\nonumber
&K_{\mu\nu\rho}
=\frac{1}{12}W^{\alpha}\varepsilon_{\alpha\mu\nu\rho}
+\frac{1}{3}\left(V_{\mu}g_{\rho\nu}-V_{\nu}g_{\rho\mu}\right)+\\
&+[K_{\mu\nu\rho}-\frac{1}{12}W^{\alpha}\varepsilon_{\alpha\mu\nu\rho}
-\frac{1}{3}\left(V_{\mu}g_{\rho\nu}-V_{\nu}g_{\rho\mu}\right)]
\end{eqnarray}
so that we may define the tensors
\begin{eqnarray}
&T_{\rho\mu\nu}=Q_{\rho\mu\nu}-\frac{1}{6}W^{\alpha}\varepsilon_{\alpha\rho\mu\nu}
-\frac{1}{3}\left(V_{\nu}g_{\rho\mu}-V_{\mu}g_{\rho\nu}\right)
\end{eqnarray}
and also
\begin{eqnarray}
&C_{\mu\nu\rho}=K_{\mu\nu\rho}
-\frac{1}{12}W^{\alpha}\varepsilon_{\alpha\mu\nu\rho}
-\frac{1}{3}\left(V_{\mu}g_{\rho\nu}-V_{\nu}g_{\rho\mu}\right)
\end{eqnarray}
as the non-completely antisymmetric traceless parts with which we have that
\begin{eqnarray}
&Q_{\rho\mu\nu}
=\frac{1}{6}W^{\alpha}\varepsilon_{\alpha\rho\mu\nu}
+\frac{1}{3}\left(V_{\nu}g_{\rho\mu}-V_{\mu}g_{\rho\nu}\right)+T_{\rho\mu\nu}
\end{eqnarray}
and also
\begin{eqnarray}
&K_{\mu\nu\rho}
=\frac{1}{12}W^{\alpha}\varepsilon_{\alpha\mu\nu\rho}
+\frac{1}{3}\left(V_{\mu}g_{\rho\nu}-V_{\nu}g_{\rho\mu}\right)+C_{\mu\nu\rho}
\end{eqnarray}
are the decompositions of torsion and contorsion in their three irreducible parts identically. 

Another notable geometrical attribute is the Levi--Civita covariant derivative $\nabla_{\mu}$ defined via the Levi--Civita symmetric connection $\Lambda^{\mu}_{\alpha\sigma}$; its associated curvature is the Riemann metric curvature tensor $R^{\rho}_{\phantom{\rho}\eta\mu\nu}$, with contraction $R_{\eta\nu}$ contracted as $R$ called Ricci metric curvature tensor and scalar as usual. The importance of the Levi--Civita torsionless connection is that with it, and the contorsion tensor, we have that the most general dynamical connection can be decomposed according $\Gamma^{\mu}_{\alpha\sigma}=\Lambda^{\mu}_{\alpha\sigma}+K^{\mu}_{\phantom{\mu}\alpha\sigma}$ which will be used to separate contorsion, or equivalently torsion, from all remaining purely metric quantities.

Equivalently, we may employ the vierbein or tetrad formalism, in which orthonormal tetrad fields $\xi_{a}^{\sigma}$ are introduced so that $g_{\alpha\beta}\xi_{a}^{\alpha}\xi_{b}^{\beta}= \eta_{ab}$, where $\eta_{ab}$ is the Minkowskian matrix, while the covariant derivative denoted with $D_{\mu}$ is defined through the dynamical spin-connection $\Gamma^{i}_{j\mu}$: conditions $D\xi=0$ and $D\eta=0$ are imposed, and when the former condition $D\xi=0$ is explicitly written it reads $\Gamma^{b}_{\phantom{b}j\mu}=\xi^{\alpha}_{j}\xi_{\rho}^{b} (\Gamma^{\rho}_{\phantom{\rho}\alpha\mu}+\xi_{\alpha}^{k}\partial_{\mu}\xi^{\rho}_{k})$ while the latter result into $\Gamma^{bj}_{\phantom{bj}\mu}=-\Gamma^{jb}_{\phantom{jb}\mu}$ so that the first formula is used to establish a link between the connection and the spin-connection while the last antisymmetry represents a property of the spin-connection that will be important to describe the underlying Lorentz structures when we shall be dealing with Lorentz spinorial representations, in the following.

In this tetrad formalism the curvature tensor is translated as
\begin{eqnarray}
&G^{a}_{\phantom{a}b\sigma\pi}
=\partial_{\sigma}\Gamma^{a}_{b\pi}-\partial_{\pi}\Gamma^{a}_{b\sigma}
+\Gamma^{a}_{j\sigma}\Gamma^{j}_{b\pi}-\Gamma^{a}_{j\pi}\Gamma^{j}_{b\sigma}
\label{Riemanngauge}
\end{eqnarray}
while the Cartan torsion tensor is 
\begin{eqnarray}
&-Q^{a}_{\phantom{a}\mu\nu}=\partial_{\mu}\xi^{a}_{\nu}-\partial_{\nu}\xi^{a}_{\mu}
+\Gamma^{a}_{j\mu}\xi^{j}_{\nu}-\Gamma^{a}_{j\nu}\xi^{j}_{\mu}
\label{Cartangauge}
\end{eqnarray}
as it is easy to check: notice that one of the advantages of this formalism is to highlight the similarities between curvature and torsion as strengths of the spin-connection and tetrad-field potentials as a Poincar\'{e} gauge theory in the Yang-Mills sense \cite{h-h-k-n}.

Before proceeding, we recall that it is possible to define a geometry of complex fields, where the gauge-covariant derivative $D_{\mu}$ is defined through the gauge-connection $A_{\mu}$.

From the gauge connection we can define
\begin{eqnarray}
&F_{\mu\nu}=\partial_{\mu}A_{\nu}-\partial_{\nu}A_{\mu}
\label{Maxwellgauge}
\end{eqnarray}
as the strength of the gauge-connection potential as an internal $U(1)$ gauge theory in the Yang-Mills sense, as it is usually done in any introduction of quantum field theory.

Another of the advantages over the coordinate formalism of the tetrad formalism is that the transformation laws are no longer general coordinate transformations but Lorentz transformations, which have an explicit structure suitable to receive other representations such as the complex representation in which the inclusion of the complex fields above would actually fit perfectly: to see how, we consider first of all the introduction of the set of complex matrices $\boldsymbol{\gamma}_{a}$ of the Clifford complex algebra $\{\boldsymbol{\gamma}_{i},\boldsymbol{\gamma}_{j}\}=2\boldsymbol{\mathbb{I}}\eta_{ij}$ giving the set of complex matrices $\boldsymbol{\sigma}_{ab}$ as $[\boldsymbol{\gamma}_{i},\boldsymbol{\gamma}_{j}]=4\boldsymbol{\sigma}_{ij}$ such that $\{\boldsymbol{\gamma}_{i},\boldsymbol{\sigma}_{jk}\}=i\varepsilon_{ijkq}\boldsymbol{\gamma}\boldsymbol{\gamma}^{q}$ and since in $(1+3)$-dimensional spacetimes all representations are equivalent, we only need to choose one and in the following we will always work in chiral representation; the set of $\boldsymbol{\sigma}_{ij}$ matrices is the set of complex generators of the infinitesimal Lorentz complex transformation, called spinorial transformation, for which the spinorial covariant derivative $\boldsymbol{D}_{\mu}$ is defined by the spinorial connection $\boldsymbol{\Gamma}_{\mu}$: the spinorial constancy of the $\boldsymbol{\gamma}_{j}$ and $\boldsymbol{\sigma}_{ij}$ matrices can be equivalently translated into the fact that the spinorial connection can be written as a Lorentz-valued connection plus an abelian term which can therefore be identified with the gauge-connection potential as $\boldsymbol{\Gamma}_{\mu}=
\frac{1}{2}\Gamma^{ab}_{\phantom{ab}\mu}\boldsymbol{\sigma}_{ab}+iqA_{\mu}\boldsymbol{\mathbb{I}}$ for any value of the $q$ label. For a list of topological properties for such global constructions, see for instance references \cite{h-m-m-n,FF}.

Now to see what these field equations actually are, we will employ the usual method of establishing on the basis of general principles a given Lagrangian and varying it with respect to the independent fields: the formalism through which a general Lagrangian depending on curvature and torsion as well as spin-connection and tetrad-fields has been established and intensively discussed in most of its consequences already in \cite{CVB,VC,VCB,CV} and we are not going to discuss it here; the point we need to underline is that this formalism is the gravitational equivalent of what would be a formalism in which the general Lagrangian depends on the field strength as well as the gauge-connection potentials, as in the usual Yang-Mills theory. Consequently, by establishing the proper matter Lagrangian, its variation with respect to matter fields will give the corresponding matter field equations.

One of the advantages of both the method presented in references \cite{h-h-k-n,h-m-m-n} and the formalism developed in references \cite{CVB,VC,VCB,CV} is that according to these there is an equivalent role covered by both curvature and torsion in the underlying framework upon which the dynamics is built, and consequently a similar equivalent importance of curvature and torsion should be expected in the gravitational dynamics as well; therefore it should not be surprising that such a complementarity between curvature and torsion should be found in the construction of the gravitational Lagrangian of the theory. However this is not the case in general, because as in the simplest HECSK theory, the field equations come from the HE Lagrangian ${\cal L}_G=G$ clearly favouring curvature over torsion; consequently within the field equations torsion enters with the same coupling constant of the curvature, that is the gravitational coupling constant, and this is the reason for which torsional effects are negligible for all practical purposes. At this point a reasonable question we may want to ask is whether the HECSK theory should not be regarded as a theory that is too simple.

Inasmuch as both curvature and torsion are taken as independent geometrical fields, a more general treatment allowing torsion not only implicitly within the curvature but also explicitly on its own would appear to be more elegant: an example of theories in which torsion must be considered explicitly and not only implicitly through the curvature, so that the Lagrangian is constructed on an intertwined mix between torsion and curvature tensors, are the conformal theories of gravity \cite{Fabbri:2011vk}; however, these theories have higher-order derivative Lagrangians, and consequently their torsion propagates in regions in which the effects of torsion are expected to be vanishingly small \cite{k-r-t}, and if torsion does actually vanish, so that the field equations of Weyl gravity are recovered \cite{Fabbri:2011gt}, there are nevertheless solutions that do not reduce to those of the known Einstein limit \cite{Fabbri:2008vp, Fabbri:2011zz} and therefore it is not surprising that discrepancies with observations do indeed arise \cite{Flanagan:2006ra}. This behaviour is not a peculiarity of conformal gravity, but of all higher-order differential theories of gravitation as a whole; a wise choice might consequently be to look for theories that have least-order derivative Lagrangians, so to recover the proper approximation. And among all of these theories, it is widely known that the HECSK theory is the simplest we have.

So we have to face a delicate situation, in which we have to carefully find a balance between the push-forward, toward a generalization in which torsion is included beside curvature into the Lagrangian, and a pull-back, for which least-order derivative Lagrangians alone are considered in order to ensure the existence of the non-propagation and therefore the expected limit. Thus we may now ask: is there a more general HECSK-like theory we can consider? An answer has been recently given in \cite{f-v}, leaving however the possibility for another question: what is the most general HECSK-like theory we can possibly have?

In order to look for this answer, we have to notice that a Lagrangian that is least-order derivative can only be linear in the curvature and quadratic in torsion: as an easy computation shows, for a linear expression of curvature we can construct only one independent scalar, the Ricci scalar $G$, and for a quadratic expression of torsion we can construct three independent scalars, given by the $Q^{\nu}_{\phantom{\nu}\nu\mu}Q_{\rho}^{\phantom{\rho}\rho\mu}$, $Q_{\rho\mu\nu}Q^{\rho\mu\nu}$, $Q_{\rho\mu\nu}Q^{\nu\mu\rho}$; the decomposition of the curvature in terms of the metric curvature plus torsion and the subsequent decomposition of torsion into its three irreducible parts shows that we may equivalently construct the Ricci metric curvature scalar $R$, with $V_{\mu}V^{\mu}$, $W_{\rho}W^{\rho}$, $T_{\rho\mu\nu}T^{\rho\mu\nu}$ proving that all irreducible fields have been accounted. Therefore, a Lagrangian that will take into account the curvature scalar and the three torsional terms will correspondingly have to be settled in terms of four coupling constants, that is the Newton gravitational coupling constant and the three torsional coupling constants given by $A$, $B$, $C$ independent on one another and yet to be determined; such a Lagrangian will therefore be written in the form 
\begin{eqnarray}
&{\cal L}_G=G+AQ^{\nu}_{\phantom{\nu}\nu\mu}Q_{\rho}^{\phantom{\rho}\rho\mu}
+BQ_{\rho\mu\nu}Q^{\rho\mu\nu}+CQ_{\rho\mu\nu}Q^{\nu\mu\rho}
\label{lagrangian}
\end{eqnarray}
as the most general that is linear in the curvature and quadratic in torsion, that is the most general least-order derivative Lagrangian defining the most general HECSK-like theory.

This gravitational Lagrangian is supplemented by the electrodynamic Lagrangian given in terms of the charge $q$ as the coupling constant of the Maxwell theory; such a Lagrangian
\begin{eqnarray}
&{\cal L}_E=-\frac{1}{4}F_{\alpha\beta}F^{\alpha\beta}
\end{eqnarray}
is the only least-order derivative Lagrangian that we could possibly define.

Finally the matter Lagrangian ${\cal L}_M$ in terms of the mass $m$ of the matter field and with Planck constant $\hbar$ normalized to the unity will complete the Lagrangian of the theory.

This complete Lagrangian, upon variation with respect to all independent field involved, will yield the field equations
\begin{eqnarray}
\nonumber
&C\left(Q^{\mu\nu\rho}-Q^{\nu\mu\rho}+2Q^{\rho\mu\nu}\right)
+B\left(2Q^{\nu\mu\rho}-2Q^{\mu\nu\rho}\right)
+A\left(V^{\nu}g^{\rho\mu}-V^{\mu}g^{\rho\nu}\right)+\\
&+\left(Q^{\rho\mu\nu}+V^{\mu}g^{\rho\nu}-V^{\nu}g^{\rho\mu}\right)=-S^{\rho\mu\nu}
\label{Cartan-Sciama-Kibble}
\end{eqnarray}
for the torsion-spin coupling and
\begin{eqnarray}
\nonumber
&C\left(D_{\mu}Q^{\mu\rho\alpha}-D_{\mu}Q^{\rho\mu\alpha}
+V_{\mu}Q^{\mu\rho\alpha}-V_{\mu}Q^{\rho\mu\alpha} +Q^{\theta\sigma\alpha}Q_{\sigma\theta}^{\phantom{\sigma\theta}\rho}
-\frac{1}{2}Q^{\theta\sigma\pi}Q_{\pi\sigma\theta}g^{\rho\alpha}\right)+\\
\nonumber
&+B\left(2D_{\mu}Q^{\alpha\rho\mu}+2V_{\mu}Q^{\alpha\rho\mu}
+2Q^{\theta\sigma\alpha}Q_{\theta\sigma}^{\phantom{\theta\sigma}\rho}
-Q^{\rho\theta\sigma}Q^{\alpha}_{\phantom{\alpha}\theta\sigma}
-\frac{1}{2}Q^{\theta\sigma\pi}Q_{\theta\sigma\pi}g^{\rho\alpha}\right)+\\
\nonumber
&+A\left(-D^{\alpha}V^{\rho}+D_{\mu}V^{\mu}g^{\rho\alpha}
+\frac{1}{2}V_{\mu}V^{\mu}g^{\rho\alpha}\right)+\\
&+\left(G^{\rho\alpha}-\frac{1}{2}Gg^{\rho\alpha}\right)
-\frac{1}{2}\left(\frac{1}{4}g^{\rho\alpha}F^{2}-F^{\rho\theta}F^{\alpha}_{\phantom{\alpha}\theta}\right)
=\frac{1}{2}T^{\rho\alpha}
\label{Hilbert-Einstein}
\end{eqnarray}
for the curvature-energy coupling, together with
\begin{eqnarray}
&\frac{1}{2}F_{\mu\nu}Q^{\rho\mu\nu}+F^{\mu\rho}Q_{\mu}+D_{\sigma}F^{\sigma\rho}=qJ^{\rho}
\label{Maxwell}
\end{eqnarray}
for the gauge-current coupling; the spin and energy densities $S^{\rho\mu\nu}$ and $T^{\rho\alpha}$ and the current density $J^{\rho}$ will be given along with the matter field equations when the Lagrangian of matter will be set explicitly: the Bianchi identities are converted by these field equations into the conservation laws given by the following expressions
\begin{eqnarray}
&D_{\rho}S^{\rho\mu\nu}+V_{\rho}S^{\rho\mu\nu}
+\frac{1}{2}T^{[\mu\nu]}=0
\label{conservationlawspin}
\end{eqnarray}
and
\begin{eqnarray}
&D_{\mu}T^{\mu\rho}+V_{\mu}T^{\mu\rho}-T_{\mu\sigma}Q^{\sigma\mu\rho}
+S_{\beta\mu\sigma}G^{\sigma\mu\beta\rho}
+qJ_{\beta}F^{\beta\rho}=0
\label{conservationlawenergy}
\end{eqnarray}
and also
\begin{eqnarray}
D_{\rho}J^{\rho}+V_{\rho}J^{\rho}=0
\label{conservationlawcurrent}
\end{eqnarray}
as conservation laws for the spin and energy and for the current density, which will turn to be valid once the matter field equations will be implemented within the model.

Before going forward, we must spend some words to study the general structure of the field equation for the torsion-spin coupling \eqref{Cartan-Sciama-Kibble} for which, as torsion or contorsion can be decomposed in three irreducible parts, the field equations can correspondingly be decomposed in three irreducible parts: then we have one field equation for the trace vector
\begin{eqnarray}
&aV_{\nu}=-S^{\rho}_{\phantom{\rho}\rho\nu}
\end{eqnarray}
one field equation for the completely antisymmetric irreducible tensor
\begin{eqnarray}
&bW_{\alpha}=-S^{\rho\mu\nu}\varepsilon_{\rho\mu\nu\alpha}
\end{eqnarray}
and one field equation for the non-completely antisymmetric irreducible tensor
\begin{eqnarray}
&cT^{\rho\mu\nu}=-\left[S^{\rho\mu\nu}
-\frac{1}{6}S^{\sigma\pi\tau}\varepsilon_{\sigma\pi\tau\alpha}\varepsilon^{\alpha\rho\mu\nu}
-\frac{1}{3}\left(S_{\sigma}^{\phantom{\sigma}\sigma\nu}g^{\rho\mu}-S_{\sigma}^{\phantom{\sigma}\sigma\mu}g^{\rho\nu}\right)\right]
\end{eqnarray}
and after renaming $3A+2B+C-2=a$, $4C-4B+1=b$, $2B+C+1=c$ as new coupling constants for the sake of simplicity, we have several possible scenarios: 
\begin{description}
\item[1. The most general torsional case:] in the case $a\neq0,\ b\neq0,\ c\neq0$ the model is the most general HECSK gravitation capable of accounting for all the parts of the spin of the matter field; this theory would be able to describe the most general higher-spin boson or fermion with no constraint whatsoever.

\item[2. Double-part decomposition of torsion:] in either of the cases $a=0,\ b\neq0,\ c\neq0$ or $b=0,\ a\neq0,\ c\neq0$ or $c=0,\ a\neq0,\ b\neq0$ the model reduces to a simpler HECSK gravitation without the possibility to couple, respectively, to the trace or the completely antisymmetric or the non-completely antisymmetric traceless part of the spin of the matter field; this theory would be able to describe more general higher-spin bosons or fermions although constraints over the spin are found also in this case, except for special instances in which particular symmetries are implemented, as for the Friedmann-Lema\^{\i}tre-Robertson-Walker spacetime, where only the two vectorial parts of torsion are compatible with cosmological constraints, so that the case given by the choice $a=0,\ b\neq0,\ c\neq0$, that is the first instance, fits perfectly.

\item[3. Single-part decomposition of torsion:] in either of the cases $b=c=0,\ a\neq0$ or $a=c=0,\ b\neq0$ or $a=b=0,\ c\neq0$ the model reduces to a simpler HECSK gravitation coupling, respectively, to the trace or the completely antisymmetric or the non-completely antisymmetric irreducible part of the spin of a given field; this theory would be able to include also more vector fields with mass and higher-spin tensor fields and spinor fields can be described, but apart from the second instance, $a=c=0,\ b\neq0$, coupling to the simplest spinor field, the Dirac fermion, in more general, higher-spin bosons or fermions, constraints over the spin are found.

\item[4. The simplest torsionless case:] in the case $a=b=c=0$ the model reduces to the simplest HECSK gravitation in which there is no coupling for any decomposition of the spin; this theory describes only fields whose spin density tensor $S^{\rho\mu\nu}$ is equal to zero, and therefore it is able to account for scalars or vector fields coming from a gauge invariance and nothing more, so that more general vector fields with mass or higher-spin tensor fields or even spinorial fields are excluded from this scheme.
\end{description}
For a thorough discussion about the decompositions of torsion see \cite{Capozziello:2001mq}. Having discussed how torsion should be decomposed, it is easy to see that in all cases $4$, $3$, $2$ the decomposition of torsion that corresponds to the vanishing coefficient cannot be determined unambiguously, and it is only in the case $1$ that it is possible to unambiguously determine all parts of torsion and invert the field equations for the torsion-spin coupling \eqref{Cartan-Sciama-Kibble} in order to get the expression of torsion in terms of the spin; then in the remaining field equations \eqref{Hilbert-Einstein} and \eqref{Maxwell} it is possible to decompose all covariant derivatives and curvatures into the Levi--Civita purely metric covariant derivatives and curvature plus contorsional contributions, which can be written in terms of torsion and eventually substituted through the field equations above in terms of the spin density of the matter fields.

In the case in which the spin density vanishes, then torsion and contorsion vanish as well, the gravitational and electrodynamic field equations are the HE and Maxwell field equations so that the Newtonian limit is recovered as desired.
%%%%%%%%%%%%%%%%%%%%%%%%%%%%%%%%%%%%%%%%%%%%%%%%%%%%%%%%%%%%%%%%%%%%%%%%%%%%%%%%%%%%%%%%%%%%%%%%%%%
\section{The coupling to fermionic fields}
In this section we apply the presented theory to a couple of fermionic fields given by spinors with spin-$\frac{1}{2}$: the special case of a Majorana-like spinor with higher-order derivative field equations, the ELKO field; and the more general spinor but with least-order derivative field equations, the usual Dirac field. For them the spin and energy densities together with the current will be given explicitly, and the matter field equations will be considered.

\subsection{The ELKO field}
A few years ago, a new form of matter field called ELKO matter field has been introduced in the panorama of physics: this form of matter field is represented by spinors with spin-$\frac{1}{2}$ $\lambda$ verifying the charge conjugation conditions $\boldsymbol{\gamma}^{2}\lambda^{*}= \pm\lambda$ for self- and antiself-conjugated fields respectively, and thus they are a type of Majorana matter field \cite{a-g/1}; with this definition, ELKOs are clearly topologically neutral fermions, and because a topological charge is what keeps localized the otherwise extended field, it follows that the absence of such charges ensures that nothing protects the field from spreading, allowing them to display non-local properties: as a consequence, according to the Wigner prescription, for which fundamental fields are classified in terms of irreducible representations of the Poincar\'{e} group, ELKOs belong to a non-standard Wigner class \cite{a-g/2}. Thus defined, one has that for the ELKO matter field to possess mass, without the introduction of Grassmann-valued fields, they have to obey second-order derivative matter field equations, and therefore they must have mass dimension $1$ \cite{a-l-s/1,a-l-s/2}; however, still referring to the Wigner classification scheme, fundamental fields are labelled in terms of both mass and spin quantum numbers, implying that matter fields have both energy and spin density tensors, and thus requiring both curvature and torsion, if ELKOs have to include the most general coupling to gravitation \cite{b/1,b/11}. All of the properties found in the ELKO matter field definition contribute to some effect in their dynamics: for instance, the ELKO matter field is proven to have $q=0$, that is it has no coupling to electrodynamics and so it is a natural candidate for dark matter \cite{a}, while their completeness relationships indicate that there must be a preferred axis of locality, probably having something to do with the privileged direction that arises within the scheme of Very Special Relativity \cite{a-h}; applications to cosmology, in particular dark energy, have been addressed in different works \cite{b/2,b-m,b-b/1,b-b/2,b-b-m-s,s/1,s/2}; their algebraic structures have been discussed quite thoroughly \cite{r-r,dr-hs,r-h,hs-dr}, whereas their dynamics in terms of exotic properties has been studied as well \cite{d-dc-hds,dr-b-hds,dr-hds,Bernardini:2012sc}. An alternative approach starting from quantum field theory formalism has been pursued \cite{g-m,Gillard:2011mv}, while a variety of phenomenological applications have been investigated \cite{w-d,l,we,ba-bh,l-z-y-c,Sadjadi:2011uu}.

Now, that ELKOs have second-order differential field equations and a coupling to torsion has two important consequences: first, the ELKO field equations have kinetic term with two derivatives, and secondly, their spin density tensor is differentially related to the torsion tensor; these two consequences taken together imply that, within the ELKO field equations, there is the appearance of derivatives of torsion, which is itself containing derivatives of the ELKOs, and thus additional second-order derivatives of the ELKOs do arise, which may induce acausal propagation and in the high-energy ultraviolet limit singularities may form. Nevertheless, all torsional fermionic back-reactions are proven to cancel exactly, ensuring the ELKO causal propagation while in the high-energy limit the ELKO gravitational pull tends to vanish, showing that ELKO matter fields are causal and they have a gravitational asymptotic freedom for which their topological non-locality is extended to include a dynamical form of gravitational non-locality \cite{fabbri/1,fabbri/2,fabbri/3}. ELKO matter has then been generalized, both in terms of the ELKO matter field dynamics itself and in terms of the coupling to the gravitational background in various ways \cite{fabbri,fabbri-vignolo,FABBRI}.

The matter fields defined as ELKO and ELKO dual are explicitly given as
\begin{eqnarray}
&\boldsymbol{\gamma}^{2}\lambda_{\pm\mp}^{*}=\eta\lambda_{\pm\mp}\ \ \ \ \ \ \ \ 
\stackrel{\neg}{\lambda}_{\pm\mp}^{*}\boldsymbol{\gamma}^{2}=-\eta\stackrel{\neg}{\lambda}_{\pm\mp}
\end{eqnarray}
with relationship between ELKO and ELKO as $\stackrel{\neg}{\lambda}_{\mp\pm}=\pm i\lambda_{\pm\mp}^{\dagger}\boldsymbol{\gamma}^{0}$ in which $\eta=\pm1$ for self- or antiself-conjugate fields and with the label $\pm\mp$ indicating that the fields decompose into irreducible chiral projections that are eigenstates with positive/negative or negative/positive eigenvalues of the helicity operator: ELKO and ELKO dual have explicit decomposition in terms of the irreducible chiral projections given by the following
\begin{eqnarray}
&\stackrel{\neg}{\lambda}_{\pm\mp}=\left(\begin{array}{cc}\pm i\eta L_{\mp}^{T}\boldsymbol{\sigma}^{2} \ \ \ \ \mp iL_{\mp}^{\dagger}
\end{array}\right)\ \ \ \ \ \ \ \ 
\begin{tabular}{c}
$\lambda_{\pm\mp}=\left(\begin{array}{c}L_{\pm}\\ -\eta \boldsymbol{\sigma}^{2}L_{\pm}^{*}
\end{array}\right)$
\end{tabular}
\end{eqnarray}
with the label $\pm$ designating the eigenstate with positive or negative eigenvalue of the helicity operator respectively given by
\begin{eqnarray}
&\stackrel{\neg}{\lambda}_{+-}=\left(\begin{array}{cc}
-\eta q\ \ 0\ \ 0\ \ -iq^{*}
\end{array}\!\right)\ \ \ \ 
\begin{tabular}{c}
$\lambda_{+-}=\left(\begin{array}{c}
d\\
0\\
0\\ 
-i\eta d^{*}
\end{array}\!\!\right)$
\end{tabular}\\
&\stackrel{\neg}{\lambda}_{-+}=\left(\begin{array}{cc}
0\ \ -\eta q\ \ iq^{*}\ \ 0
\end{array}\!\right)\ \ \ \
\begin{tabular}{c}
$\lambda_{-+}=\left(\begin{array}{c}
0\\
d\\
i\eta d^{*}\\ 
0 
\end{array}\!\!\right)$
\end{tabular}
\end{eqnarray}
when their spin and momentum have the same direction; all ELKO special algebraic properties come from their special structure in terms of Majorana-like spinors, while there is nothing special in their differential properties as these depend only on the fact that they are spin-$\frac{1}{2}$ spinor fields. Given the specific algebraic structure and the usual differential structure we may now proceed and write the most general ELKO Lagrangian.

The Lagrangian we will employ will be the most general we can possibly write, and since ELKO fields are spinors of mass dimension $1$ this consequently means, first of all, that the Lagrangian for ELKO must account for all possible terms built with two derivatives, and finally, that for ELKO there are two such terms: the first is the one we would have have in the simplest case $\boldsymbol{D}_{\alpha}\stackrel{\neg}{\lambda}\boldsymbol{D}^{\alpha}\lambda$ but also the term $\boldsymbol{D}_{\alpha}\stackrel{\neg}{\lambda}\boldsymbol{\sigma}^{\alpha\beta} \boldsymbol{D}_{\beta}\lambda$ can be taken since due to the presence of torsion it does not reduce to the divergence of a vector, negligible in the action; therefore, the most general of such Lagrangians for ELKO will eventually be given with both these terms, and one normalization factor aside, it will contain one parameter $p$ to be determined. The most complete Lagrangian for the ELKO field is then given by the following form
\begin{eqnarray}
{\cal L}_{\textrm{ELKO}}=-\boldsymbol{D}_{\alpha}\stackrel{\neg}{\lambda}(\eta^{\alpha\beta}\mathbb{I} +p\boldsymbol{\sigma}^{\alpha\beta})\boldsymbol{D}_{\beta}\lambda+m^{2}\stackrel{\neg}{\lambda}\lambda
\end{eqnarray}
in terms of the mass $m$ of the matter field.

Its variation gives the spin and energy densities
\begin{eqnarray}
\nonumber
&S_{\mu\alpha\beta}
=\frac{1}{2}\left(\boldsymbol{D}_{\mu}\stackrel{\neg}{\lambda}
\boldsymbol{\sigma}_{\alpha\beta}\lambda
-\stackrel{\neg}{\lambda}\boldsymbol{\sigma}_{\alpha\beta}
\boldsymbol{D}_{\mu}\lambda\right)+\\
&+\frac{p}{2}\left(\boldsymbol{D}^{\rho}\stackrel{\neg}{\lambda}
\boldsymbol{\sigma}_{\rho\mu}\boldsymbol{\sigma}_{\alpha\beta}\lambda
-\stackrel{\neg}{\lambda}\boldsymbol{\sigma}_{\alpha\beta}
\boldsymbol{\sigma}_{\mu\rho}\boldsymbol{D}^{\rho}\lambda\right)
\label{spin}
\end{eqnarray}
and
\begin{eqnarray}
\nonumber
&T_{\mu\nu}=
\left(\boldsymbol{D}_{\mu}\stackrel{\neg}{\lambda}\boldsymbol{D}_{\nu}\lambda
+\boldsymbol{D}_{\nu}\stackrel{\neg}{\lambda}\boldsymbol{D}_{\mu}\lambda
-g_{\mu\nu}\boldsymbol{D}_{\rho}\stackrel{\neg}{\lambda}\boldsymbol{D}^{\rho}\lambda\right)+\\
&+p\left(\boldsymbol{D}_{\nu}\stackrel{\neg}{\lambda}
\boldsymbol{\sigma}_{\mu\rho}\boldsymbol{D}^{\rho}\lambda
+\boldsymbol{D}^{\rho}\stackrel{\neg}{\lambda}
\boldsymbol{\sigma}_{\rho\mu}\boldsymbol{D}_{\nu}\lambda
-g_{\mu\nu}\boldsymbol{D}_{\rho}\stackrel{\neg}{\lambda}
\boldsymbol{\sigma}^{\rho\sigma}\boldsymbol{D}_{\sigma}\lambda\right)
+g_{\mu\nu}m^{2}\stackrel{\neg}{\lambda}\lambda
\label{energy}
\end{eqnarray}
without field equation for the current density since ELKO fields are neutral and while the matter field equations are
\begin{eqnarray}
&\left(\boldsymbol{D}^{2}\lambda+V^{\mu}\boldsymbol{D}_{\mu}\lambda\right)
+p\left(\boldsymbol{\sigma}^{\rho\mu}\boldsymbol{D}_{\rho}\boldsymbol{D}_{\mu}\lambda
+V_{\rho}\boldsymbol{\sigma}^{\rho\mu}\boldsymbol{D}_{\mu}\lambda\right)
+m^{2}\lambda=0
\end{eqnarray}
to be taken into account: as soon as the matter field equations are employed the energy and spin densities undergo the conservation laws \eqref{conservationlawspin} and \eqref{conservationlawenergy} above.

The whole system of field equations is thus given by the field equation for the torsion
\begin{eqnarray}
\nonumber
&-2C\left(Q^{\mu\nu\rho}-Q^{\nu\mu\rho}+2Q^{\rho\mu\nu}\right)
-2B\left(2Q^{\nu\mu\rho}-2Q^{\mu\nu\rho}\right)
-2A\left(V^{\nu}g^{\rho\mu}-V^{\mu}g^{\rho\nu}\right)+\\
\nonumber
&-2\left(Q^{\rho\mu\nu}+V^{\mu}g^{\rho\nu}-V^{\nu}g^{\rho\mu}\right)
=\left(\boldsymbol{D}^{\rho}\stackrel{\neg}{\lambda}\boldsymbol{\sigma}^{\mu\nu}\lambda
-\stackrel{\neg}{\lambda}\boldsymbol{\sigma}^{\mu\nu}\boldsymbol{D}^{\rho}\lambda\right)+\\
&+p\left(\boldsymbol{D}_{\pi}\stackrel{\neg}{\lambda}\boldsymbol{\sigma}^{\pi\rho}
\boldsymbol{\sigma}^{\mu\nu}\lambda-\stackrel{\neg}{\lambda}\boldsymbol{\sigma}^{\mu\nu}
\boldsymbol{\sigma}^{\rho\pi}\boldsymbol{D}_{\pi}\lambda\right)
\end{eqnarray}
and the curvature
\begin{eqnarray}
\nonumber
&2C\left(D_{\mu}Q^{\mu\rho\alpha}-D_{\mu}Q^{\rho\mu\alpha}
+V_{\mu}Q^{\mu\rho\alpha}-V_{\mu}Q^{\rho\mu\alpha} +Q^{\theta\sigma\alpha}Q_{\sigma\theta}^{\phantom{\sigma\theta}\rho}
-\frac{1}{2}Q^{\theta\sigma\pi}Q_{\pi\sigma\theta}g^{\rho\alpha}\right)+\\
\nonumber
&+2B\left(2D_{\mu}Q^{\alpha\rho\mu}+2V_{\mu}Q^{\alpha\rho\mu}
+2Q^{\theta\sigma\alpha}Q_{\theta\sigma}^{\phantom{\theta\sigma}\rho}
-Q^{\rho\theta\sigma}Q^{\alpha}_{\phantom{\alpha}\theta\sigma}
-\frac{1}{2}Q^{\theta\sigma\pi}Q_{\theta\sigma\pi}g^{\rho\alpha}\right)+\\
\nonumber
&+2A\left(-D^{\alpha}V^{\rho}+D_{\mu}V^{\mu}g^{\rho\alpha}
+\frac{1}{2}V_{\mu}V^{\mu}g^{\rho\alpha}\right)+\\
\nonumber
&+\left(2G^{\rho\alpha}-Gg^{\rho\alpha}\right)
=\left(\boldsymbol{D}^{\rho}\stackrel{\neg}{\lambda}\boldsymbol{D}^{\alpha}\lambda
+\boldsymbol{D}^{\alpha}\stackrel{\neg}{\lambda}\boldsymbol{D}^{\rho}\lambda
-g^{\rho\alpha}\boldsymbol{D}_{\pi}\stackrel{\neg}{\lambda}\boldsymbol{D}^{\pi}\lambda\right)+\\
&+p\left(\boldsymbol{D}^{\alpha}\stackrel{\neg}{\lambda}
\boldsymbol{\sigma}^{\rho\pi}\boldsymbol{D}_{\pi}\lambda
+\boldsymbol{D}_{\pi}\stackrel{\neg}{\lambda}
\boldsymbol{\sigma}^{\pi\rho}\boldsymbol{D}^{\alpha}\lambda
-g^{\rho\alpha}\boldsymbol{D}_{\pi}\stackrel{\neg}{\lambda}
\boldsymbol{\sigma}^{\pi\sigma}\boldsymbol{D}_{\sigma}\lambda\right)
+g^{\rho\alpha}m^{2}\stackrel{\neg}{\lambda}\lambda
\end{eqnarray}
along with the matter field equations
\begin{eqnarray}
&\left(\boldsymbol{D}^{2}\lambda+V^{\mu}\boldsymbol{D}_{\mu}\lambda\right)
+p\left(\boldsymbol{\sigma}^{\rho\mu}\boldsymbol{D}_{\rho}\boldsymbol{D}_{\mu}\lambda
+V_{\rho}\boldsymbol{\sigma}^{\rho\mu}\boldsymbol{D}_{\mu}\lambda\right)
+m^{2}\lambda=0
\end{eqnarray}
which are rather complicated, and the decomposition of the torsion-spin coupling field equations, let alone the decomposition of torsion and its substitution with the spin in all other field equations, may be very difficult to perform. 

However, we may employ the freedom this theory gives us to choose special fine-tunings for all parameters involved, which considerably simplify the solution of this problem, and the following treatment; this is due to the fact that in such cases, in which some of the parameters $a$, $b$, $c$ vanish, the corresponding trace, completely antisymmetric and non-completely antisymmetric traceless decompositions of torsion are not uniquely defined, and they may therefore be chosen to vanish without any loss of generality. In the next sections we are going to show how these simplification will be performed in two special examples in cosmological applications of particular interest.

\paragraph{1. The FLRW-compatible ELKO field with $\boldsymbol{i\stackrel{\neg}{\lambda} \boldsymbol{\gamma}\lambda\equiv0}$ constraint.} To study cosmological applications, we have to implement the cosmological principle, for which the metric is Friedmann-Lema\^{\i}tre-Robertson-Walker metric, which will be chosen to be spatially flat as
\begin{eqnarray}
&g_{tt}=1,\ \ \ \ \ \ \ \ g_{jj}=-\sigma^{2}\ \ \ \ j=x,y,z
\end{eqnarray}
where torsion has only the time component of the vector parts $V_{t}$ and $W_{t}$; the ELKO field that turns out to be compatible with such symmetries may be chosen to be given by
\begin{eqnarray}
&\stackrel{\neg}{\lambda}_{+-}=\sqrt{2}\varphi\left(\begin{array}{cc}
\eta\ \ 0\ \ 0\ \ i
\end{array}\!\right)\ \ \ \ 
\begin{tabular}{c}
$\lambda_{+-}=\sqrt{2}\varphi\left(\begin{array}{c}
1\\
0\\
0\\ 
-i\eta
\end{array}\!\!\right)$
\end{tabular}\\
&\stackrel{\neg}{\lambda}_{-+}=\sqrt{2}\varphi\left(\begin{array}{cc}
0\ \ \eta\ \ -i\ \ 0
\end{array}\!\right)\ \ \ \
\begin{tabular}{c}
$\lambda_{-+}=\sqrt{2}\varphi\left(\begin{array}{c}
0\\
1\\
i\eta\\ 
0 
\end{array}\!\!\right)$
\end{tabular}
\end{eqnarray}
with the additional property for which $i\stackrel{\neg}{\lambda}\boldsymbol{\gamma}\lambda\equiv0$ identically: in this way torsion has only the time component of the vector part $V_{t}$ if we assume $p=-1$ for the ELKO matter.

In this case it is possible to see that the field equations for the torsion-spin coupling are decomposed to the single independent field equation for the torsion-spin trace vector
\begin{eqnarray}
&aV_{\nu}=\partial_{\nu}\left(\frac{3\lambda^{2}}{8}\right)
\end{eqnarray}
so that the field equations for the curvature-energy coupling are
\begin{eqnarray}
\nonumber
&-2\left(\frac{2+a}{3}\right)\left(D^{\alpha}V^{\rho}-D_{\mu}V^{\mu}g^{\rho\alpha}
-\frac{1}{2}V_{\mu}V^{\mu}g^{\rho\alpha}\right)+\\
\nonumber
&+\left(2G^{\rho\alpha}-Gg^{\rho\alpha}\right)
=\left(\boldsymbol{D}^{\rho}\stackrel{\neg}{\lambda}\boldsymbol{D}^{\alpha}\lambda
+\boldsymbol{D}^{\alpha}\stackrel{\neg}{\lambda}\boldsymbol{D}^{\rho}\lambda
-g^{\rho\alpha}\boldsymbol{D}_{\pi}\stackrel{\neg}{\lambda}\boldsymbol{D}^{\pi}\lambda\right)+\\
&-\left(\boldsymbol{D}^{\alpha}\stackrel{\neg}{\lambda}
\boldsymbol{\sigma}^{\rho\pi}\boldsymbol{D}_{\pi}\lambda
+\boldsymbol{D}_{\pi}\stackrel{\neg}{\lambda}
\boldsymbol{\sigma}^{\pi\rho}\boldsymbol{D}^{\alpha}\lambda
-g^{\rho\alpha}\boldsymbol{D}_{\pi}\stackrel{\neg}{\lambda}
\boldsymbol{\sigma}^{\pi\sigma}\boldsymbol{D}_{\sigma}\lambda\right)
+g^{\rho\alpha}m^{2}\lambda^{2}
\end{eqnarray}
and we have now to separate torsion away everywhere; after this is done, the field equations for the Ricci tensor are
\begin{eqnarray}
\nonumber
&2R^{\rho\alpha}=\frac{2a}{3}\left(\nabla^{\alpha}V^{\rho}
+\frac{1}{2}\nabla_{\mu}V^{\mu}g^{\rho\alpha}
+\frac{1}{3}V^{\rho}V^{\alpha}-\frac{1}{3}V^{\mu}V_{\mu}g^{\rho\alpha}\right)
+\frac{1}{4}V^{\rho}\nabla^{\alpha}\lambda^{2}+\\
\nonumber
&+\frac{1}{3}V_{\beta}\left(\boldsymbol{\nabla}^{\rho}\stackrel{\neg}{\lambda}
\boldsymbol{\sigma}^{\beta\alpha}\lambda
-\stackrel{\neg}{\lambda}\boldsymbol{\sigma}^{\beta\alpha}\boldsymbol{\nabla}^{\rho}\lambda
+\stackrel{\neg}{\lambda}\boldsymbol{\sigma}^{\beta\alpha}
\boldsymbol{\sigma}^{\rho\pi}\boldsymbol{\nabla}_{\pi}\lambda
-\boldsymbol{\nabla}_{\pi}\stackrel{\neg}{\lambda}\boldsymbol{\sigma}^{\pi\rho}
\boldsymbol{\sigma}^{\beta\alpha}\lambda\right)+\\
&+\left(\boldsymbol{\nabla}^{\rho}\stackrel{\neg}{\lambda}\boldsymbol{\nabla}^{\alpha}\lambda
+\boldsymbol{\nabla}^{\alpha}\stackrel{\neg}{\lambda}\boldsymbol{\nabla}^{\rho}\lambda
-\boldsymbol{\nabla}^{\alpha}\stackrel{\neg}{\lambda}
\boldsymbol{\sigma}^{\rho\pi}\boldsymbol{\nabla}_{\pi}\lambda
-\boldsymbol{\nabla}_{\pi}\stackrel{\neg}{\lambda}
\boldsymbol{\sigma}^{\pi\rho}\boldsymbol{\nabla}^{\alpha}\lambda\right)
-g^{\rho\alpha}m^{2}\lambda^{2}
\end{eqnarray}
in which now torsion has to be substituted in terms of the ELKO fields: and after this is done as well, the time-time component will be such that only one term will contain the parameter $a$ so that for $a$ very small this term will be much larger than all others leaving
\begin{eqnarray}
&\frac{\ddot{\sigma}}{\sigma}\approx-\frac{1}{a}\left(\varphi\dot{\varphi}\right)^{2}
\end{eqnarray}
describing the evolution of the scale factor $\sigma$ as a function of the cosmological time.

As it is easy to see for $a$ that approaches zero from negative values we have that the acceleration $\ddot{\sigma}$ is positive and very high, therefore this may turn out to give relevant contributions for the accelerated expansion of the universe.

\paragraph{2. The FLRW-compatible ELKO field with $\boldsymbol{\stackrel{\neg}{\lambda} \lambda\equiv0}$ constraint.} We continue the study of cosmological application by still considering the Friedmann-Lema\^{\i}tre-Robertson-Walker spatially flat metric
\begin{eqnarray}
&g_{00}=1,\ \ \ \ \ \ \ \ g_{jj}=-\sigma^{2}\ \ \ \ j=x,y,z
\end{eqnarray}
where torsion has only the time component of the vector parts $V_{t}$ and $W_{t}$; however in this case the ELKO field will be chosen to be
\begin{eqnarray}
&\stackrel{\neg}{\lambda}_{+-}=\sqrt{2}\varphi\left(\begin{array}{cc}
i\eta\ \ 0\ \ 0\ \ 1
\end{array}\!\right)\ \ \ \ 
\begin{tabular}{c}
$\lambda_{+-}=\sqrt{2}\varphi\left(\begin{array}{c}
1\\
0\\
0\\ 
-i\eta
\end{array}\!\!\right)$
\end{tabular}\\
&\stackrel{\neg}{\lambda}_{-+}=\sqrt{2}\varphi\left(\begin{array}{cc}
0\ \ i\eta \ \ -1\ \ 0
\end{array}\!\right)\ \ \ \
\begin{tabular}{c}
$\lambda_{-+}=\sqrt{2}\varphi\left(\begin{array}{c}
0\\
1\\
i\eta\\ 
0 
\end{array}\!\!\right)$
\end{tabular}
\end{eqnarray}
with the additional property for which $\lambda^{2}\equiv0$ identically: in this way torsion has only the time component of the vector part $W_{t}$ if we assume $p=-1$ for the ELKO matter.

The field equations for the torsion-spin coupling are decomposed to the one independent field equation for the torsion-spin completely antisymmetric dual of the axial vector
\begin{eqnarray}
&bW_{\nu}=\partial_{\nu}\left(\frac{3i\stackrel{\neg}{\lambda}\boldsymbol{\gamma}\lambda}{4}\right)
\end{eqnarray}
so that the field equations for the curvature-energy coupling are
\begin{eqnarray}
\nonumber
&\left(\frac{1-b}{6}\right)\left(D_{\mu}W_{\theta}\varepsilon^{\mu\theta\rho\alpha}
+\frac{1}{6}W^{\alpha}W^{\rho}+\frac{1}{12}W^{2}g^{\rho\alpha}\right)+\\
\nonumber
&+\left(2G^{\rho\alpha}-Gg^{\rho\alpha}\right)
=\left(\boldsymbol{D}^{\rho}\stackrel{\neg}{\lambda}\boldsymbol{D}^{\alpha}\lambda
+\boldsymbol{D}^{\alpha}\stackrel{\neg}{\lambda}\boldsymbol{D}^{\rho}\lambda
-g^{\rho\alpha}\boldsymbol{D}_{\pi}\stackrel{\neg}{\lambda}\boldsymbol{D}^{\pi}\lambda\right)+\\
&-\left(\boldsymbol{D}^{\alpha}\stackrel{\neg}{\lambda}
\boldsymbol{\sigma}^{\rho\pi}\boldsymbol{D}_{\pi}\lambda
+\boldsymbol{D}_{\pi}\stackrel{\neg}{\lambda}
\boldsymbol{\sigma}^{\pi\rho}\boldsymbol{D}^{\alpha}\lambda
-g^{\rho\alpha}\boldsymbol{D}_{\pi}\stackrel{\neg}{\lambda}
\boldsymbol{\sigma}^{\pi\sigma}\boldsymbol{D}_{\sigma}\lambda\right)
\end{eqnarray}
and we have now to separate torsion away everywhere; after this is done, the field equations for the Ricci tensor are
\begin{eqnarray}
\nonumber
&2R^{\rho\alpha}=\frac{b}{6}\left(\nabla_{\mu}W_{\theta}\varepsilon^{\mu\theta\rho\alpha}
+\frac{1}{6}W^{\alpha}W^{\rho}-\frac{1}{6}W^{2}g^{\rho\alpha}\right)
+\frac{1}{16}W^{\rho}\nabla^{\alpha}(i\stackrel{\neg}{\lambda}\boldsymbol{\gamma}\lambda)+\\
\nonumber
&+\frac{i}{12}W_{\beta}\left(\boldsymbol{\nabla}^{\rho}\stackrel{\neg}{\lambda}
\boldsymbol{\sigma}^{\beta\alpha}\boldsymbol{\gamma}\lambda
-\stackrel{\neg}{\lambda}\boldsymbol{\gamma}
\boldsymbol{\sigma}^{\beta\alpha}\boldsymbol{\nabla}^{\rho}\lambda
+\stackrel{\neg}{\lambda}\boldsymbol{\gamma}\boldsymbol{\sigma}^{\beta\alpha}
\boldsymbol{\sigma}^{\rho\pi}\boldsymbol{\nabla}_{\pi}\lambda
-\boldsymbol{\nabla}_{\pi}\stackrel{\neg}{\lambda}\boldsymbol{\sigma}^{\pi\rho}
\boldsymbol{\sigma}^{\beta\alpha}\boldsymbol{\gamma}\lambda\right)+\\
&+\left(\boldsymbol{\nabla}^{\rho}\stackrel{\neg}{\lambda}\boldsymbol{\nabla}^{\alpha}\lambda
+\boldsymbol{\nabla}^{\alpha}\stackrel{\neg}{\lambda}\boldsymbol{\nabla}^{\rho}\lambda
-\boldsymbol{\nabla}^{\alpha}\stackrel{\neg}{\lambda}
\boldsymbol{\sigma}^{\rho\pi}\boldsymbol{\nabla}_{\pi}\lambda
-\boldsymbol{\nabla}_{\pi}\stackrel{\neg}{\lambda}
\boldsymbol{\sigma}^{\pi\rho}\boldsymbol{\nabla}^{\alpha}\lambda\right)
\end{eqnarray}
in which now torsion has to be substituted in terms of the ELKO fields: and after this is done as well, the time-time component will be such that only one term will contain the parameter $b$ so that for $b$ very small this term will be much larger than all others leaving
\begin{eqnarray}
&\frac{\ddot{\sigma}}{\sigma}\approx-\frac{1}{2b}\left(\varphi\dot{\varphi}\right)^{2}
\end{eqnarray}
describing the evolution of the scale factor $\sigma$ as a function of the cosmological time.

And for $b$ that approaches zero from negative values the acceleration $\ddot{\sigma}$ is positive and very high, turning out to be relevant for the accelerated expansion of the universe.

It is intriguing that in these two complementary examples we eventually get an analogous result at the end: of the four contributions to the ELKO energy, whether they are the standard or the additional one proportional to the parameter $p$, whether in terms of the torsionless derivative and the torsional potentials, the one coming from the additional energy proportional to the parameter $p$ in the torsional potentials, that is the one that has the most genuine torsional origin, is the only one that remains to be relevant when the remaining parameter is chosen to approach zero from negative values; once this is sorted out, it controls the behaviour of the accelerated expansion of the universe.
%%%%%%%%%%%%%%%%%%%%%%%%%%%%%%%%%%%%%%%%%%%%%%%%%%%%%%%%%%%%%%%%%%%%%%%%%%%%%%%%%%%%%%%%%%%%%%%%%%%
\subsection{The Dirac field}
One of the reasons for which torsion was long neglected is to be found in the historical misunderstanding that the implementation of the principles of equivalence and causality was supposed to force the symmetry of the connection and consequently implying the vanishing of the torsion itself \cite{w,m-t-w}; this misunderstanding however may be resolved by a deeper analysis, and if care is taken it is straightforward to acknowledge that even in the case in which the principle of equivalence holds then torsion is only restricted to be completely antisymmetric \cite{m-l,xy,so,Fabbri:2006xq,Fabbri:2009se}; it is therefore quite intriguing that if torsion is completely antisymmetric then the spin must be completely antisymmetric as well, and the only spin that is completely antisymmetric without any constraint imposed is that of the simplest spinor field \cite{Fabbri:2008rq,Fabbri:2009yc}; this simplest spinor field is indeed the Dirac field, for which the completely antisymmetric dual of an axial current appearing as a self-interaction in the Dirac field equation is able to give a dynamical explanation of the exclusion principle as stated by Pauli \cite{Fabbri:2010rw}. The study of the Dirac equation is for this reason quite interesting.

The Dirac field Lagrangian is notoriously the only least-order derivative Lagrangian
\begin{eqnarray}
{\cal L}_{\textrm{Dirac}}
=\frac{i}{2}(\bar{\psi}\boldsymbol{\gamma}^{\mu}\boldsymbol{D}_{\mu}\psi
-\boldsymbol{D}_{\mu}\bar{\psi}\boldsymbol{\gamma}^{\mu}\psi)
-m\bar{\psi}\psi
\end{eqnarray}
in terms of the mass $m$ of the matter field.

Its variation gives the completely antisymmetric spin and energy densities
\begin{eqnarray}
&S_{\mu\alpha\beta}=
\frac{i}{4}\bar{\psi}\{\boldsymbol{\gamma}_{\mu},\boldsymbol{\sigma}_{\alpha\beta}\}\psi
\end{eqnarray}
and
\begin{eqnarray}
&T_{\mu\alpha}=\frac{i}{2}(\bar{\psi}\boldsymbol{\gamma}_{\mu}\boldsymbol{D}_{\alpha}\psi-\boldsymbol{D}_{\alpha}\bar{\psi}\boldsymbol{\gamma}_{\mu}\psi)
\end{eqnarray}
and the current
\begin{eqnarray}
&J_{\mu}=\bar{\psi}\boldsymbol{\gamma}_{\mu}\psi
\end{eqnarray}
as soon as the matter field equations
\begin{equation}
\label{fematter}
i\boldsymbol{\gamma}^{\mu}\boldsymbol{D}_{\mu}\psi
+\frac{i}{2}\boldsymbol{\gamma}^{\mu}V_{\mu}\psi-m\psi=0
\end{equation}
are taken into account, and where the matter field equations also imply that the energy and spin densities and the current obey the conservation laws \eqref{conservationlawspin} and \eqref{conservationlawenergy} above.

The whole system of field equations is thus given by the field equation for the completely antisymmetric torsion
\begin{eqnarray}
&bW^{\sigma}=\frac{3}{2}\bar{\psi}\boldsymbol{\gamma}^{\sigma}\boldsymbol{\gamma}\psi
\end{eqnarray}
and the curvature
\begin{eqnarray}
\nonumber
&\left(\frac{1-b}{12}\right)\left(D_{\mu}W_{\sigma}\varepsilon^{\mu\sigma\rho\alpha}
+\frac{1}{6}W^{\alpha}W^{\rho}+\frac{1}{12}W^{2}g^{\alpha\rho}\right)
+\left(G^{\rho\alpha}-\frac{1}{2}Gg^{\rho\alpha}\right)-\\
&-\frac{1}{2}\left(\frac{1}{4}g^{\rho\alpha}F^{2}-F^{\rho\theta}F^{\alpha}_{\phantom{\alpha}\theta}\right)
=\frac{i}{4}(\bar{\psi}\boldsymbol{\gamma}^{\rho}\boldsymbol{D}^{\alpha}\psi
-\boldsymbol{D}^{\alpha}\bar{\psi}\boldsymbol{\gamma}^{\rho}\psi)
\end{eqnarray}
and also the gauge field
\begin{eqnarray}
&\frac{1}{12}F_{\mu\nu}W_{\sigma}\varepsilon^{\mu\nu\sigma\rho}+D_{\sigma}F^{\sigma\rho}
=q\bar{\psi}\boldsymbol{\gamma}^{\rho}\psi
\end{eqnarray}
along with the matter field equations
\begin{equation}
i\boldsymbol{\gamma}^{\mu}\boldsymbol{D}_{\mu}\psi-m\psi=0
\end{equation}
for which the conservation laws \eqref{conservationlawspin} and \eqref{conservationlawenergy} above will turn out to be important for the decomposition in torsionless and torsional quantities within all the field equations.

After such a separation and upon substitution with the spin density of the Dirac field, we get the field equations for the Ricci tensor
\begin{eqnarray}
&R^{\rho\alpha}-\frac{1}{2}\left(\frac{1}{4}g^{\rho\alpha}F^{2}-F^{\rho\theta}F^{\alpha}_{\phantom{\alpha}\theta}\right)
=\frac{i}{8}\left(\bar{\psi}\boldsymbol{\gamma}^{(\rho}\boldsymbol{\nabla}^{\alpha)}\psi
-\boldsymbol{\nabla}^{(\alpha}\bar{\psi}\boldsymbol{\gamma}^{\rho)}\psi\right)
-\frac{1}{4}g^{\rho\alpha}m\psi^{2}
\end{eqnarray}
and the gauge field
\begin{eqnarray}
&\nabla_{\sigma}F^{\sigma\rho}=q\bar{\psi}\boldsymbol{\gamma}^{\rho}\psi
\end{eqnarray}
along with the matter field equations
\begin{eqnarray}
&i\boldsymbol{\gamma}^{\mu}\boldsymbol{\nabla}_{\mu}\psi
+\frac{3}{16b}\bar{\psi}\boldsymbol{\gamma}^{\mu}\boldsymbol{\gamma}\psi
\boldsymbol{\gamma}_{\mu}\boldsymbol{\gamma}\psi-m\psi=0
\end{eqnarray}
showing that the gravitational field equations decomposed into their symmetric part written in terms of the Ricci tensor and the electrodynamic field equations are formally those we would have in the torsionless case while the matter field equations become formally identical to what we would have had in the torsionless case with additional potentials of self-interactions having an independent coupling constant, and where the gravitational field equations in their antisymmetric part are implied by the spin conservation law \eqref{conservationlawspin}.

Focusing on the non-gravitational field equations, we see that they may be written after a Fierz rearrangement as
\begin{eqnarray}
&\nabla_{\sigma}F^{\sigma\rho}=q\bar{\psi}\boldsymbol{\gamma}^{\rho}\psi
\end{eqnarray}
together with
\begin{eqnarray}
&i\boldsymbol{\gamma}^{\mu}\boldsymbol{\nabla}_{\mu}\psi
-\frac{3}{16b}\bar{\psi}\boldsymbol{\gamma}^{\mu}\psi\boldsymbol{\gamma}_{\mu}\psi-m\psi=0
\end{eqnarray}
showing that the vector $\bar{\psi}\boldsymbol{\gamma}^{\mu}\psi$ accounts for both electrodynamic and self-interaction.

A further simplification may be given in terms of another Fierz rearrangement while having the coupling constant renamed as $\lambda=\frac{3}{16b}$ to get
\begin{eqnarray}
&i\boldsymbol{\gamma}^{\mu}\boldsymbol{\nabla}_{\mu}\psi
-\lambda\left(\bar{\psi}\psi-\bar{\psi}\boldsymbol{\gamma}\psi\boldsymbol{\gamma}\right)\psi
-m\psi=0
\end{eqnarray}
as the matter field equations will now study in specific situations.

We remark that in the present theory as in the HECSK theory, the Dirac field equations are of the Nambu-Jona--Lasinio type \cite{Fabbri:2010rw}; however, in this model the Dirac field has self-interactions whose coupling constant is not determined \cite{Fabbri:2011kq}. It is important to notice that no matter what this value is, it would at first be thinkable to normalize it to the usual value of the Newton constant through the normalization of the Dirac field, but a deeper analysis shows that this process would also have the effect of changing the scale of the energy density within the gravitational field equations and it is thus unacceptable.

So far as our knowledge is concerned, the theory we have here constructed is the most general in which the HECSK theory for the Dirac field can be generalized in order for the system of field equations, once decomposed in terms of torsionless quantities plus torsional contributions written as Dirac field self-interactions, to provide a free coupling constant for the self-interactions of matter while leaving the gravitation unmodified. 

\subsubsection{The Dirac field self-coupling and its applications:\\ Leptons with weak interactions and Higgs field} To begin with, we take into account the case of a coupled system of Dirac spinorial matter fields, one of which is a spinor with mass while the other is a semispinor and therefore only compatible with the massless configuration; these spinors will possess spin-torsion coupling giving interactions between the two fundamental fields while in the standard model the couple of electron and neutrino may only have weak interactions: the aim is therefore to compare this case with that of the weak forces and see whether or not similarities arise. 

In this case the field equations for the electron and neutrino fields $e$ and $\nu$ are given by the previous field equations, but because now the system is constituted by a couple of spinors then the spin will be the sum of the two spins and the spinorial field equations are
\begin{eqnarray}
&i\boldsymbol{\gamma}^{\mu}\boldsymbol{\nabla}_{\mu}e
-\lambda\left(\overline{e}\boldsymbol{\gamma}_{\mu}e\boldsymbol{\gamma}^{\mu}e
+\overline{\nu}\boldsymbol{\gamma}_{\mu}\nu\boldsymbol{\gamma}^{\mu}\boldsymbol{\gamma}e\right)-me=0\\
&i\boldsymbol{\gamma}^{\mu}\boldsymbol{\nabla}_{\mu}\nu
-\lambda\overline{e}\boldsymbol{\gamma}_{\mu}\boldsymbol{\gamma}e\boldsymbol{\gamma}^{\mu}\nu=0
\end{eqnarray}
in which the fact that the electron is massive while the neutrino is massless is the reason that prevents these two fields to mix into a doublet; equivalently we have
\begin{eqnarray}
\label{equations}
\nonumber
&i\boldsymbol{\gamma}^{\mu}\boldsymbol{\nabla}_{\mu}e
+2\lambda(\cos{\theta})^{2}\overline{e}\boldsymbol{\gamma}e\boldsymbol{\gamma}e+\\
&+q\tan{\theta}Z_{\mu}\boldsymbol{\gamma}^{\mu}e
-\frac{g}{2\cos{\theta}}Z_{\mu}\boldsymbol{\gamma}^{\mu}e_{L}
+\frac{g}{\sqrt{2}}W^{*}_{\mu}\boldsymbol{\gamma}^{\mu}\nu-He-me=0
\label{electron}\\
&i\boldsymbol{\gamma}^{\mu}\boldsymbol{\nabla}_{\mu}\nu
+\frac{g}{2\cos{\theta}}Z_{\mu}\boldsymbol{\gamma}^{\mu}\nu
+\frac{g}{\sqrt{2}}W_{\mu}\boldsymbol{\gamma}^{\mu}e_{L}=0
\label{neutrino}
\end{eqnarray}
once we define
\begin{eqnarray}
&Z^{\mu}
=-\lambda\left[2(\sin{\theta})^{2}\overline{e}\boldsymbol{\gamma}^{\mu}e
-\overline{e}_{L}\boldsymbol{\gamma}^{\mu}e_{L}
+\overline{\nu}\boldsymbol{\gamma}^{\mu}\nu\right]\left(\frac{\cot{\theta}}{q}\right)
\label{neutral}\\
&W^{\mu}
=-\lambda\left(\overline{e}_{L}\boldsymbol{\gamma}^{\mu}\nu\right)
\left[\frac{4(\sin{\theta})^{2}-1}{q\sqrt{2}\sin{\theta}}\right]
\label{charged}\\
&H=\lambda\overline{e}e2(\cos{\theta})^{2}
\label{Higgs}
\end{eqnarray}
showing that field equations (\ref{electron}) and (\ref{neutrino}) are formally identical to the system of field equations for the lepton fields after the symmetry breaking in the standard model, although both weak and Higgs boson fields are here composite \cite{Fabbri:2009ta}. This is an important discrepancy with respect to the standard model, because, in our approach, we expect both weak and Higgs bosons to display internal structure whenever the energy is high enough to probe their potential compositeness, while, in the standard model, they are supposed to be structureless at any energy scale. Similar results are obtained for leptons in Higgsless models and therefore without symmetry breaking \cite{f/1}; for leptons and hadrons in circumstances in which the symmetry was unbroken see references \cite{f/2,f/3}. The phenomenon of oscillations for neutrino in the massless configuration have been studied in reference \cite{F/1}.

In order to reproduce the weak interactions the tuning $\lambda=G_{F}$ ensures the coupling constant to have the value of the Fermi coupling constant while in the case of oscillations the constant $\lambda$ must be tuned to the oscillation length: the field equations that have to be solved in order to get the relationship between $\lambda$ and the oscillation length are non-linear and therefore numerical analysis must be used. But even though it is unlikely that the experimental value for $\lambda$ in this case may fit that of the Fermi model.

\subsubsection{The Dirac field self-coupling and its applications:\\ the Dirac field modelling condensed state physics} We will now write the Dirac field equation for the single field in the standard representation, the one that allows the slow-speed weak-field limit, in which the spinor field maintains only the large component given by the semispinor $\phi$, and for which the spinor field equation is split into two field equations of which one reduces to zero while the other is the field equation for the semispinor $\phi$ given by
\begin{eqnarray}
&i\frac{\partial\phi}{\partial t}
+\frac{1}{2m}\boldsymbol{\sigma}^{k}\boldsymbol{\nabla}_{k}
\boldsymbol{\sigma}^{a}\boldsymbol{\nabla}_{a}\phi
-\lambda\left(\phi^{\dagger}\phi\right)\phi-m\phi=0
\label{matterfieldapproximated}
\end{eqnarray}
compatibly with the fact that the spinor actually has only two degrees of freedom, and representing cold-matter fields with self-interactions with the structure of a mass term.

Thus in the present theory we have that the Dirac field equations have the form of the Nambu-Jona--Lasinio field equation further approximated to the Pauli-Schr\"{o}dinger field equation of the Ginzburg-Landau type \cite{Fabbri:2010rw}. In it the coupling constant $\lambda$ is to be set on the value of the specific condensed state system we would eventually like to study \cite{Fabbri:2011kq}.

From now on we will neglect the electrodynamic field, which in particular means that the magnetic field will be vanishing, thus permitting the decoupling between the spin degrees of freedom, so that the semispinor will correspondingly decouple in the two components given by the scalars $u$ and $v$, and after the field equations have been separated as well, the $u$ and $v$ field will have to obey the coupled system of field equations
\begin{eqnarray}
&i\frac{\partial u}{\partial t}
+\frac{1}{2m}\nabla^{2}u
-\lambda u^{2}u-(m+\lambda v^{2})u=0\\
&i\frac{\partial v}{\partial t}
+\frac{1}{2m}\nabla^{2}v
-\lambda v^{2}v-(m+\lambda u^{2})v=0
\label{matterfieldapproximatedreduced}
\end{eqnarray}
representing cold-matter fields after a process of bosonization, with self-interactions and interactions between the two components; notice that in the approximation for which for one field the other field's square may be considered constant, then the residual interaction becomes a correction to the mass term, and the two equations may be treated separately.

Thus in the present theory we have that the field equations are approximated to the Pauli-Schr\"{o}dinger field equation of the Ginzburg-Landau type and then decoupled to the Schr\"{o}dinger field equation of the Gross-Pitaevskii type. In it the coupling constant $\lambda$ has to be set according to the value of the specific condensed state system we would eventually like to study and as it is clear it may be attractive as well as repulsive as we shall see.

If we now focus on the field $u$ alone, it is possible to see that for stationary states the field equation reduces to
\begin{eqnarray}
&\frac{1}{2m}\nabla^{2}u-\lambda u^{2}u+\epsilon u=0
\end{eqnarray}
with total energy $\epsilon$ either positive or negative, and whose general solution may be given according to the sign of the constants and parameters involved, therefore we have a total of four different cases: for attractive potentials we will set $m\lambda=-g^{2}$ while for repulsive potentials we will set $m\lambda=g^{2}$; and in each case, negative energy will be written in terms of $2m\epsilon=-\omega^{2}$ and positive energy will be written in terms of $2m\epsilon=\omega^{2}$ in what follows.

Now an interesting application is that of the unidimensional potentials in the coordinate $x$, as in this case exact solutions are easy to find: for attractive potentials the energy may only be negative and a solution is given by
\begin{eqnarray}
&u(x)=\frac{\omega}{g}\frac{1}{\cosh{(\omega x)}}
\end{eqnarray}
whereas for repulsive potentials we still have the case of negative energy with solution
\begin{eqnarray}
&u(x)=\frac{\omega}{g}\frac{1}{\sinh{(\omega x)}}
\end{eqnarray}
and that of positive energy with solution
\begin{eqnarray}
&u(x)=\frac{\omega}{g}\frac{1}{\sin{(\omega x)}}
\label{spin-chain}
\end{eqnarray}
as it can be checked by direct substitution; it is easy to see that while for attractive potentials we always have bound states that are finite and tend to zero at infinity, for repulsive potentials the bound states diverge as $\frac{1}{x}$ but still tend to zero at infinity whereas the unbound states diverge as $\frac{1}{x}$ and exhibit a periodic behaviour. If we focus on repulsive potentials we have that alternative solutions are obtained for negative energy as
\begin{eqnarray}
&u(x)=\frac{\omega}{\sqrt{2}g}\tan{\left(\frac{\omega x}{\sqrt{2}}\right)}
\end{eqnarray}
and for positive energy as
\begin{eqnarray}
&u(x)=\frac{\omega}{\sqrt{2}g}\tanh{\left(\frac{\omega x}{\sqrt{2}}\right)}
\label{soliton}
\end{eqnarray}
again as it can be checked directly; in this case there is the tendency to spread at infinity, since what is supposed to be the bound states exhibit a periodic behaviour while the unbound states display the solitonic behaviour. The repulsive potential with positive energy is interesting since a general solution can be found in terms of the Jacobi elliptic functions, and because torsion should have the behaviour of an angular momentum giving rise to potentials with the structure of centrifugal barriers and the energy is usually assumed to be positive, and therefore it represents the most reasonable physical situation.

However, for a complete study of the unidimensional potential, suitable boundary conditions must be assumed, and therefore we will next turn our attention to the unidimensional potential box with width $l$: we focus on the case of repulsive potential with positive energy supposing that the repulsive potential is much smaller than the energy so that $g$ is much smaller than $\omega$; an approximated solution is given by
\begin{eqnarray}
&u_{n}(x)=\sin{(\omega_{n}x)}
+\frac{g^{2}}{\omega^{2}}\left[\frac{1}{16}\sin{(3\omega_{n}x)}
-\frac{3}{4}\omega_{n}x\cos{(\omega_{n}x)}\right]
\label{particle}
\end{eqnarray}
with boundary condition $u(0)=u(l)=0$ giving discrete $\omega_{n}$ and thus energy levels $\epsilon_{n}$ as
\begin{eqnarray}
&\epsilon_{n}=n^{2}\frac{\pi^{2}}{2ml^{2}}+\lambda\frac{3}{4}
\end{eqnarray}
which is therefore clearly quantized, as expected for any bound state with such a set of boundary conditions; notice that even in the ground-state configuration the torsional interaction generates a zero-point energy that is not equal to zero. This is quite intriguing, and further calculations should be performed to get higher orders of approximation.

Unidimensional plane waves may also be employed to evaluate the general form of the energy levels for the free system to be
\begin{eqnarray}
&\epsilon=m+\frac{p^{2}}{2m}+\lambda u^{2}
\end{eqnarray}
with $u^{2}$ being the density of the field itself, which in the interpretation of a collective state of plane waves represents the number of waves per unit volume $u^{2}=\frac{N}{V}$; on the other hand, it is also possible to go a little further and employ a thermodynamic interpretation of the kinetic energy as the temperature $T$ of the system: considering both interpretations, the energy levels are reinterpreted as
\begin{eqnarray}
&\epsilon=T+\lambda \frac{N}{V}
\label{energyvdw}
\end{eqnarray} 
which is the expression for the energy level of the Van der Waals gas. By employing the equation of the energy $\left(\frac{\partial \epsilon}{\partial V}\right)_{T}= \left(\frac{\partial P}{\partial T}\right)_{V}T-P$ it is easy to see that the equation of state gives the pressure $P$ as a function of volume and temperature as
\begin{eqnarray}
&P-\lambda\frac{N}{V^{2}}-\frac{T}{F(V)}=0
\end{eqnarray}
as the Van der Waals equation of state with generalized volume factor $F(V)$ and with coupling constant $\lambda$: for negative values the Van der Waals pressure gives an attraction, for positive values the Van der Waals pressure gives a repulsion. In this reinterpretation the matter field is seen as a classic wave distribution behaving as a Van der Waals gas.
%%%%%%%%%%%%%%%%%%%%%%%%%%%%%%%%%%%%%%%%%%%%%%%%%%%%%%%%%%%%%%%%%%%%%%%%%%%%%%%%%%%%%%%%%%%%%%%%%%%
\section{Conclusion}
In this paper, we have constructed a generalization of the Einstein-Cartan gravity in which both curvature and torsion are considered each with its own coupling constant, the one related to the curvature being the gravitational constant while the three related to torsion are still undetermined; we have applied this geometrical background to the case of Dirac and ELKO fields showing that the field equations are endowed with self-interactions whose coupling constants are yet to be determined: we have discussed different applications in which these self-interactions play different roles according to the value of the constant we assign, whether they are the ELKO field used in the standard model of cosmology to fit the accelerated expansion of the universe or the Dirac field both in the standard model of particle physics to fit the behaviour of fermions with weak interactions or for the condensed state approximation. We have insofar witnessed that the presence of torsion with independent coupling constants may either give rise to new forms of interactions, in cosmology, or to interactions that are eventually shown to have the same structure of already known interactions and potentials which have always been ascribed to something else, whether the context was the domain of particle physics or condensed state physics.

A problem we have already discussed in \cite{f-v} was related to the fact that the approach followed there had only one additional coupling constant which might have been unable to have the whole spectrum of possible applications fit into a single framework; the approach followed here has three coupling constants, and so freer in its structure: for instance, it is possible to consider both ELKO and Dirac field in a single Lagrangian, in a FLRW universe and choosing a specific form for the ELKO so that $i\stackrel{\neg}{\lambda}\boldsymbol{\gamma}\lambda=0$ we get that the torsion only has the trace part sourced by this ELKO field and the completely antisymmetric dual of the axial vector sourced by the Dirac field. Then it is possible to assign the values to the coupling constants as to have $c=0$ in order to ensure the compatibility with the symmetries of the background, $a$ approaching zero from negative values so that ELKO may fit the accelerated expansion of the universe and $b$ approaching zero from positive values to have Dirac fields with repulsive self-interactions able to give rise to the weak forces or bosonization in condensed states. This \emph{a posteriori} shows that it is reasonable to consider the three different parts of torsion independent as they are sourced by fields that are totally unrelated; in particular, this implies that it is reasonable to assign three different coupling constants, and employ them for different purposes. But this also shows that this freedom is anyway rather limited, as for a given field the constant is assigned once and for all, and it is unlikely that the very same field may have self-interactions with strengths that differ from one situation to another. Therefore the discussion outlined in our previous work is here improved for what regards the possibility to set different strengths for different parts of torsion, whether we are in cosmology or quantum physics, but not so much as to set different strengths for the same part of torsion in particle physics or condensed state physics, and for that an even more general theory of gravity may be able to provide a running coupling constant, possibly scaling with the energy, such that all applications above, and maybe more, can fit into a single scheme, as anticipated in \cite{FV}.

If this approach really works, torsion would not only be observable, but its effects might have already been observed, although we have not been able to recognize them as coming from torsion; and more effects may be on the verge of being discovered. In writing this we are aware of the fact that torsion may not be the answer for all problems of physics, but it possibly is for some of them, and our theory can allow us to see how.
%%%%%%%%%%%%%%%%%%%%%%%%%%%%%%%%%%%%%%%%%%%%%%%%%%%%%%%%%%%%%%%%%%%%%%%%%%%%%%%%%%%%%%%%%%%%%%%%%%%

%%%%%%%%%%%%%%%%%%%%%%%%%%%%%%%%%%%%%%%%%%%%%%%%%%%%%%%%%%%%%%%%%%%%%%%%%%%%%%%%%%%%%%%%%%%%%%%%%%%
\end{document}